\documentclass[a4paper]{article}
\usepackage{amsmath}
\usepackage{siunitx}
\usepackage{microtype}
\usepackage{INTERSPEECH2022}
\usepackage{multirow}
\usepackage[colorlinks,linkcolor=black,urlcolor=black,citecolor=black]{hyperref}

\title{On Metric Learning for Audio-Text Cross-Modal Retrieval}
\name{Xinhao Mei, Xubo Liu, Jianyuan Sun, Mark D. Plumbley, Wenwu Wang}
\address{
  Centre for Vision, Speech and Signal Processing (CVSSP)\\
        University of Surrey, UK}
\email{\{x.mei, xubo.liu, jianyuan.sun, m.plumbley, w.wang\}@surrey.ac.uk}

\begin{document}

\maketitle
\begin{abstract}
Audio-text retrieval aims at retrieving a target audio clip or caption from a pool of candidates given a query in another modality. Solving such cross-modal retrieval task is challenging because it not only requires learning robust feature representations for both modalities, but also requires capturing the fine-grained alignment between these two modalities. Existing cross-modal retrieval models are mostly optimized by metric learning objectives as both of them attempt to map data to an embedding space, where similar data are close together and dissimilar data are far apart. Unlike other cross-modal retrieval tasks such as image-text and video-text retrievals, audio-text retrieval is still an unexplored task. In this work, we aim to study the impact of different metric learning objectives on the audio-text retrieval task. We present an extensive evaluation of popular metric learning objectives on the AudioCaps and Clotho datasets. We demonstrate that NT-Xent loss adapted from self-supervised learning shows stable performance across different datasets and training settings, and outperforms the popular triplet-based losses. Our code is available at \url{https://github.com/XinhaoMei/audio-text_retrieval}.

\end{abstract}



\section{Introduction}

Given an audio clip or a caption as a query, audio-text retrieval aims at retrieving a paired item from a pool of candidates in another modality. This cross-modal retrieval task is challenging as it requires not only learning robust feature representations for both acoustic and textual modalities but also capturing fine-grained interaction between the learned acoustic and textual features and aligning them in a shared embedding space. Audio-text retrieval can be potentially applied to applications such as film, audio book production and web search. 

Cross-modal retrieval tasks (e.g. image-text retrieval and video-text retrieval) have received extensive attention in recent years and have made great progress \cite{faghri2018vse++, li2019vsrn, li2020oscar, Liu2019ce, miech2018moe, gabeur2020mmt}. However, little attention has been paid to audio-text retrieval in the literature. One reason might be the lack of appropriate datasets, thus early works just focused on tag-based audio retrieval, where the queries were words not sentences. Chechik et al. \cite{chechik2008tag_audio_retrieval} proposed a tag-based audio retrieval system using traditional machine learning techniques (e.g. support vector machines and Gaussian mixture models). Ikawa et al. \cite{ikawa2018onomatopoeic} investigated searching sounds using onomatopoeic words. Elizalde et al. \cite{elizalde2019tag_siamese} employed a siamese network to align audio and textual features to a joint embedding space. Although these tag-based audio retrieval works show reasonable performance, they are constrained in the query format. Retrieving audio clips using free-form language (sentences) is more natural for humans.

With the fast development of audio captioning in recent three years \cite{Mei2021ACT, mei2021diverse, Liu2021cl4ac, liu2022acbert}, publicly available audio captioning datasets are released \cite{kim2019audiocaps, drossos2020clotho}, which are naturally suited for the free-form language-based audio-text retrieval task. Koepke et al. \cite{koepke2022audioretrieval} first established the benchmarks for free-form language-based audio retrieval, where they adapted models from video retrieval and made use of pre-trained models to solve the data scarcity problem. Since both the audio and text (captions) are sequence data, free-form language-based audio-text retrieval is more challenging than tag-based audio retrieval and is the focus of this paper. We use the term audio-text and audio-caption interchangeably in this paper. 

Similar to other cross-modal retrieval models \cite{faghri2018vse++}, the audio-text retrieval models can be built with two sub-networks, namely, an audio encoder and a text encoder. The objectives of these two encoders are to map the audios and texts into a joint embedding space, where the semantically similar embeddings are close to each other and the dissimilar items are far away. We refer to the embeddings as Acoustic Semantic Embeddings (ASE) as they are learned via jointly modeling the audio and language modalities. The training objective of cross-modal retrieval models is consistent with that of metric learning \cite{musgrave2020metriclearning}. To this end, metric learning has been a popular choice for the optimization of the cross-modal retrieval models. Numerous metric learning objectives have been introduced for various tasks such as face identification \cite{schroff2015triplet}, speaker recognition \cite{chung2020defence}, and retrieval \cite{faghri2018vse++, li2019vsrn}, however, there is no clear argument which one is the most suited since they may work well on specific tasks or data but may not generalize well to other tasks \cite{bleeker2022lessons}. In this work, we aim to study and compare the impact of different metric learning objectives for the free-form language-based audio-text retrieval task in a constant training setting.


We focus on triplet loss and its variants, as they have demonstrated promising performance and are popularly employed \cite{schroff2015triplet}. In a triplet setting for audio-text retrieval, an audio clip and its corresponding caption are regarded as an anchor and a positive example, respectively, while other unpaired captions are regarded as negatives. The hinge-based triplet ranking loss sums over all negative samples within a mini-batch (thus we refer to it as triplet-sum). Faghri et al. \cite{faghri2018vse++} argued that hard negatives should be emphasised as other easy negatives may dominate the loss and create local minimal, thus they proposed a triplet ranking loss with hard negative mining (we refer to it as triplet-max) which focuses only on the hardest negative within a mini-batch. Wei et al. \cite{Wei_2020triplet_weight} further proposed a universal weighting framework for cross-modal retrieval, where the pairs are weighted based on their similarity scores (we refer to it as triplet-weighted). In addition to the triplet-based losses, we further adapt a contrastive loss used in self-supervised learning here for supervised cross-modal retrieval, that is, normalized temperature-scaled cross entropy loss (NT-Xent) \cite{chen2020ntxent}. The NT-Xent loss is based on softmax and aims to identify the positive pairs within a mini-batch.


In summary, we first establish a baseline using  pre-trained models to learn the acoustic semantic embeddings, then we present an extensive evaluation of the popular metric learning objectives described above on our baseline in a constant training setting. Against popular belief \cite{faghri2018vse++, bleeker2022lessons}, we demonstrate that triplet losses with hard negative mining are sensitive to the training settings and may be hard to converge, the NT-Xent loss shows stable performance with respect to different datasets and training settings and outperforms the triplet-based losses.

\section{Audio-Text Retrieval with Metric Learning}
\label{sec:objectives}
In this section, we first formulate the audio-text retrieval problem and introduce the baseline model, then the metric learning objectives (loss functions) we evaluated are introduced.

\subsection{Problem formulation}
Let $D=\{(a_i, t_i)\}_{i=1}^N$ be an audio captioning dataset of $N$ examples, where $a_i$ is an audio clip and $t_i$ is the paired caption. Therefore, $(a_i, t_i)$ is regarded as a positive pair while $(a_i, t_{j, j\neq i})$ is a negative pair. One audio clip could have multiple paired captions, we just consider one here for simplicity. The audio-text retrieval models usually consist of an audio encoder $f$ and a text encoder $g$, which project the audio clip and text into a shared embedding space, respectively. For an audio-caption pair $(a_i, t_j)$, the similarity of the audio and caption can be measured by cosine similarity of their embeddings:
\begin{equation}
    s_{ij} = \frac{f(a_i) \cdot g(t_j)}{||f(a_i)||_2||g(t_j)||_2}.
    \label{eq:sim}
\end{equation}
The two encoders are trained to make the similarity score of positive pairs $s_{ii}$ higher than that of negative pairs $s_{ij}$. 

\subsection{Model}
The data available in the audio captioning datasets is limited, and this data scarcity problem usually limits the performance of the model in learning robust feature representations. Transfer learning has been adopted as a standard recipe to alleviate the data scarcity problem and has shown promising performance in audio captioning task \cite{Mei2021ac_trans}. Therefore, pre-trained models are employed here. 

\noindent \textbf{Audio Encoder}
Pre-trained audio neural networks (PANNs) \cite{kong2020panns} are pre-trained on AudioSet with an audio tagging task, which are shown to provide robust audio representations and perform well on a wide range of audio-related tasks. The ResNet-38 in PANNs is employed as the audio encoder, where the last two linear layers are discarded. An average and max pooling layer is applied to aggregate along the frequency dimension on the feature map output by the last convolutional block. A multilayer perceptron (MLP) block is used to project the audio features into a shared embedding space, which consists of two linear layers with a ReLU \cite{glorot2011relu} activation layer between them. 

\noindent \textbf{Text Encoder} Numerous large-scale per-trained language models have been published in recent year, which show powerful capability to model the language and produce contextual-aware embeddings. BERT \cite{devlin2019bert}, which stands for Bidirectional Encoder Representations from Transformers, obtains state-of-the-art results on a wide variety of Natural Language Processing (NLP) tasks. The pre-trained BERT is employed as the text encoder here. A ``\texttt{\textless CLS\textgreater}'' token is appended at the start of each sentence and used as the final sentence representation. A MLP block is also applied to project the sentence representation into the shared embedding space.

\subsection{Loss functions}
During training, we sample a mini-batch of audio-caption pairs $\{a_i, t_i\}_{i=1}^B$ where $B$ is the batch size. Triplet-based loss functions are based on the concept of triplet, which is made up by an anchor, a positive (paired candidate in another modality) and a negative (unpaired candidate in another modality). The anchor with its positive is a positive pair and the anchor with its negative is a negative pair as we defined above.

\noindent \textbf{Triplet-sum} For each query, the triplet-sum loss aims to maximize the similarity score of its positive pair while minimizing the similarity scores to all other negatives within a mini-batch, thus it can be formulated as
\begin{equation}
    \mathcal L = \frac{1}{B} \sum_{i=1}^{B} \sum_{j\neq i}[m + s_{ij} - s_{ii}]_{+} + [m + s_{ji} - s_{ii}]_{+},
    \label{eq:triplet}
\end{equation}
where $[x]_+ = \max(0, x)$ and $m$ is a distance margin. Since audio-text retrieval is a bidirectional retrieval task (audio-to-text and text-to-audio), the loss has two terms, the first term sums over all negative captions given a query audio clip while the second term sums over all negative audio clips given a query caption. If the similarity of the positive pair is larger than that of any negatives in the mini-batch by the margin $m$, the loss will be zero. 

\noindent \textbf{Triplet-max} Triplet-sum loss sums over all negatives for each query within a mini-batch, Faghri et al. \cite{faghri2018vse++} argued that the easy negatives may dominate the loss and make it stuck into local minimal, thus, hard negatives should be emphasized. They proposed the triplet-max loss that focuses on the hardest negatives during training, which can be formulated as:
\begin{equation}
    \mathcal L = \frac{1}{B} \sum_{i=1}^{B} \max_{j\neq i}{[m + s_{ij} - s_{ii}]_{+}} + \max_{j\neq i}{[m + s_{ji} - s_{ii}]_{+}}.
    \label{eq:triplet_hard}
\end{equation}
For each query, it aims to maximize the similarity score of its positive pair while just minimizing the similarity score to its hardest negative within a mini-batch, that is, the negative closest to the query in the embedding space. In this way, the easy negatives won't violate the loss and the hardest negative takes all the gradients.

\noindent \textbf{Triplet-weighted} 
Both the positive and negative pairs are treated equally in the triplet-sum and triplet-max losses. Wei et al. \cite{Wei_2020triplet_weight} further introduced a universal function $G(\cdot)$ to weight the pairs based on their similarity scores. Specifically, the weighting function is defined as a polynomial function. For a positive pair $(a_i, t_i)$, the positive weight function $G_{\rm {pos}}$ is defined as:
\begin{equation}
    G_{\rm {pos}} = a_p s_{ii}^p + a_{p-1} s_{ii}^{p-1} + \cdots + a_1 s_{ii}^1 + a_0,
    \label{gpos}
\end{equation}
where $p$ is the order of the polynomial function and $\{a_i\}_{i=0}^{p}$ are the hyper-parameters. The negative weight function $G_{\rm {neg}}$ can be formulated as:
\begin{equation}
    G_{\rm {neg}} = b_q s_{ij}^q + b_{q-1} s_{ij}^{q-1} + \cdots + b_1 s_{ij}^1 + b_0,
    \label{gneg}
\end{equation}
where $\{b_i\}_{i=0}^{q}$ are the hyper-parameters and $q$ is the order. If the similarity score of the positive pair $(a_i, t_i)$ increases, the positive weight value will decrease. In contrast, for a negative pair $(a_i, t_j)$, the negative weight value increases if the similarity score of the negative pair increases.

\begin{table*}[!t]
  \caption{Results of the experiments. * indicates the learning rate is \num{5e-5} otherwise \num{1e-4}.}
  \label{tab:results}
  \centering
  \begin{tabular}{ccccccccc}
    \toprule
    \multirow{2}{*}{\textbf{Dataset}} & \multirow{2}{*}{\textbf{Fine-tune}} & \multirow{2}{*}{\textbf{Objective}} & \multicolumn{3}{c}{\textbf{Text-to-Audio}} & \multicolumn{3}{c}{\textbf{Audio-to-Text}}\\
    \cline{4-9}
    & & & \textbf{R$@1$} &\textbf{R$@5$} & \textbf{R$@10$} & \textbf{R$@1$} &\textbf{R$@5$} & \textbf{R$@10$}\\
    \midrule
    \multirow{8}{*}{AudioCaps} & \multirow{4}{*}{No} & Triplet-sum & $17.2_{\pm0.5}$ & $47.6_{\pm0.2}$ & $64.3_{\pm0.1}$ & $19.3_{\pm1.5}$ & $51.4_{\pm0.8}$ & $67.2_{\pm0.7}$ \\
    & & Triplet-max & $19.9_{\pm0.2}$ & $51.6_{\pm0.3}$ & $67.4_{\pm0.3}$ & $21.2_{\pm0.9}$ & $54.9_{\pm1.4}$ & $70.2_{\pm1.0}$ \\ 
    & & Triplet-weighted & $19.9_{\pm0.2}$ & $52.1_{\pm0.7}$ & $67.6_{\pm0.6}$ & $21.9_{\pm0.8}$ & $56.3_{\pm0.3}$ & $71.5_{\pm0.4}$ \\
    & & NT-Xent & $19.2_{\pm0.4}$ & $51.1_{\pm0.2}$ & $66.6_{\pm0.2}$ & $21.3_{\pm0.5}$ & $53.6_{\pm0.7}$ & $69.8_{\pm0.5}$ \\ 
    \cline{2-9}
    & \multirow{4}{*}{Yes} & Triplet-sum & $32.2_{\pm0.3}$ & $68.2_{\pm0.6}$ & $81.6_{\pm0.5}$ & $36.1_{\pm1.2}$ & $69.2_{\pm1.3}$ & $81.4_{\pm1.7}$  \\
    & & Triplet-max & $32.7_{\pm0.3}$ & $68.3_{\pm0.8}$ & $81.6_{\pm0.5}$ & $38.7_{\pm1.0}$ & $70.6_{\pm0.7}$ & $82.2_{\pm0.8}$\\
    & & Triplet-weighted & $32.6_{\pm0.6}$ & $67.7_{\pm0.7}$ & $81.0_{\pm0.8}$ & $39.6_{\pm1.1}$ & $72.0_{\pm2.2}$ & $82.2_{\pm1.4}$ \\
    & & NT-Xent & $33.9_{\pm0.4}$ & $69.7_{\pm0.2}$ & $82.6_{\pm0.3}$ & $39.4_{\pm1.0}$ & $72.0_{\pm1.0}$ & $83.9_{\pm0.6}$\\
    \midrule
    \multirow{8}{*}{Clotho} & \multirow{4}{*}{No} & Triplet-sum & $7.0_{\pm0.2}$ & $23.0_{\pm0.3}$ & $34.8_{\pm0.5}$ & $8.3_{\pm0.3}$ & $25.4_{\pm0.8}$ & $36.7_{\pm0.5}$ \\
    & & Triplet-max & $7.9_{\pm0.2}$ & $23.6_{\pm0.4}$ & $34.2_{\pm0.5}$ & $8.8_{\pm1.0}$ & $25.7_{\pm1.0}$ & $36.3_{\pm0.8}$ \\
    & & Triplet-weighted* & $7.2_{\pm0.2}$ & $22.1_{\pm0.3}$ & $33.0_{\pm0.3}$ & $7.5_{\pm0.4}$ & $23.2_{\pm0.1}$ & $33.3_{\pm0.3}$\\
    & & NT-Xent & $8.0_{\pm0.2}$ & $25.3_{\pm0.1}$ & $36.9_{\pm0.3}$ & $9.2_{\pm0.7}$ & $27.9_{\pm0.5}$ & $39.0_{\pm0.7}$\\
    \cline{2-9}
    & \multirow{4}{*}{Yes} & Triplet-sum & $14.2_{\pm0.5}$ & $36.6_{\pm0.5}$ & $49.3_{\pm0.7}$ & $16.1_{\pm0.7}$ & $37.5_{\pm1.2}$ & $50.7_{\pm1.0}$ \\
    & & Triplet-max* & $14.2_{\pm0.9}$ & $36.5_{\pm1.2}$ & $49.0_{\pm0.4}$ & $15.8_{\pm0.4}$ & $36.4_{\pm1.6}$ & $49.6_{\pm2.0}$\\
    & & Triplet-weighted* & $14.2_{\pm0.4}$ & $36.6_{\pm0.5}$ & $49.7_{\pm0.3}$ & $16.9_{\pm0.4}$ & $38.1_{\pm0.2}$ & $51.4_{\pm0.2}$ \\
    & & NT-Xent & $14.4_{\pm0.4}$ & $36.6_{\pm0.2}$ & $49.9_{\pm0.2}$ & $16.2_{\pm0.7}$ & $37.5_{\pm0.9}$ & $50.2_{\pm0.7}$ \\
    \bottomrule
  \end{tabular}
\end{table*}
Let $N_{a_i} = \{s_{ij,j\neq i}\}$ and $N_{t_j} = \{s_{ji,i\neq j}\}$ be the similarity scores for all negative pairs of an audio sample $a_i$ and a text sample $t_j$ respectively. The loss can be formulated as:
\begin{equation}
\begin{split}
    \mathcal L &= \frac{1}{B} \sum_{i=1}^{B}[\sum_{p=0}^{P}a_p s_{ii}^p + \sum_{q=0}^Q b_q\max\{N_{a_i}^q\}]_+\\
    & + \frac{1}{B} \sum_{i=1}^{B}[\sum_{p=0}^{P}a_p s_{ii}^p + \sum_{q=0}^Q b_q\max\{N_{t_i}^q\}]_+,
\end{split}
\end{equation}
where $P$ and $Q$ are the highest power of positive and negative pairs, respectively. The $\max$ term in the equation will select the hardest negative pair, thus the loss is referred to as the maximum polynomial loss. They also proposed another average polynomial loss that will first select informative negative pairs based on a mining policy and average the similarity scores of the selected pairs. We just employ the maximum polynomial loss here as it performs better than the average one in \cite{Wei_2020triplet_weight}.

\noindent \textbf{NT-Xent}
NT-Xent is a contrastive loss based on softmax, proposed by Chen et al. \cite{chen2020ntxent} to learn visual representations via self-supervised learning. We adapt it here for the supervised cross-modal retrieval task, as follows:
\begin{equation}
\begin{split}
    \mathcal L  = -\frac{1}{B} &\left ( \sum_{i=1}^{B}\log \frac{\exp({s_{ii}/\tau})}{\sum_{j=1}^B\exp{({s_{ij}/\tau)}}} + \right.\\
     & \left. \sum_{i=1}^{B}\log \frac{\exp({s_{ii}/\tau})}{\sum_{j=1}^B\exp{({s_{ji}/\tau)}}} \right),
    \label{eq:ntxent}
\end{split}
\end{equation}
where $\tau$ is a temperature hyper-parameter. It aims to maximize the similarity of the positive pair with respect to all negative pairs within a mini-batch, and the final loss is also computed in a bidirectional manner. 

\section{Experiments}

\subsection{Datasets}
We focus on free-form language-based audio-text retrieval, thus audio captioning datasets are naturally employed, in which each audio clip is annotated by humans using natural sentences. Extensive experiments are carried out on AudioCaps \cite{kim2019audiocaps} and Clotho \cite{drossos2020clotho} datasets, respectively.

\noindent \textbf{AudioCaps} AudioCaps is the largest audio captioning dataset with around \num{50}k audio clips. All the audio clips are \num{10}-second long and are sourced from AudioSet \cite{audioset}, the largest dataset for audio tagging. In our downloaded version, the training set contains \num{49274} audio clips and each of them has one corresponding human-annotated caption, the validation and test set contain \num{494} and \num{957} audio clips, respectively, and each of them has five human-annotated captions.

\noindent \textbf{Clotho} Clotho is an audio captioning dataset whose audio clips are collected from the Freesound\footnote{\url{https://freesound.org/}} archive. The length of the audio clips ranges uniformly from \num{15} to \num{30} seconds. Clotho v2 is used here. There are \num{3839} audio clips in the training set and \num{1045} audio clips in the validation and test sets respectively. All the audio clips have five diverse human-annotated captions of eight to \num{20} words in length.

\begin{table*}[t]
  \caption{Experimental results with different batch sizes on AudioCaps dataset. The pre-trained encoders are not fine-tuned.}
  \label{tab:results_batch}
  \centering
  \begin{tabular}{ccccccccc}
    \toprule
    \multirow{2}{*}{\textbf{Batch size}} & \multirow{2}{*}{\textbf{Objective}} & \multicolumn{3}{c}{\textbf{Text-to-Audio}} & \multicolumn{3}{c}{\textbf{Audio-to-Text}}\\
    \cline{3-8}
    & & \textbf{R$@1$} &\textbf{R$@5$} & \textbf{R$@10$} & \textbf{R$@1$} &\textbf{R$@5$} & \textbf{R$@10$}\\
    \midrule
    \multirow{4}{*}{32} & Triplet-sum & $17.2_{\pm0.5}$ & $47.6_{\pm0.2}$ & $64.3_{\pm0.1}$ & $19.3_{\pm1.5}$ & $51.4_{\pm0.8}$ & $67.2_{\pm0.7}$ \\
    & Triplet-max & $19.9_{\pm0.2}$ & $51.6_{\pm0.3}$ & $67.4_{\pm0.3}$ & $21.2_{\pm0.9}$ & $54.9_{\pm1.4}$ & $70.2_{\pm1.0}$ \\ 
    & Triplet-weighted & $19.9_{\pm0.2}$ & $52.1_{\pm0.7}$ & $67.6_{\pm0.6}$ & $21.9_{\pm0.8}$ & $56.3_{\pm0.3}$ & $71.5_{\pm0.4}$ \\
    & NT-Xent & $19.2_{\pm0.4}$ & $51.1_{\pm0.2}$ & $66.6_{\pm0.2}$ & $21.3_{\pm0.5}$ & $53.6_{\pm0.7}$ & $69.8_{\pm0.5}$ \\ 
    \midrule
    \multirow{4}{*}{128} & Triplet-sum & $17.0_{\pm0.7}$ & $47.2_{\pm0.2}$ & $62.7_{\pm0.2}$ & $20.0_{\pm1.2}$ & $49.9_{\pm1.5}$ & $66.7_{\pm1.0}$ \\
    & Triplet-max & $11.8_{\pm0.3}$ & $38.1_{\pm0.9}$ & $53.7_{\pm0.5}$ & $15.0_{\pm0.5}$ & $43.4_{\pm0.6}$ & $59.6_{\pm0.7}$ \\ 
    & Triplet-weighted & $10.1_{\pm0.6}$ & $31.7_{\pm0.9}$ & $46.5_{\pm0.9}$ & $10.7_{\pm0.9}$ & $33.4_{\pm1.7}$ & $48.7_{\pm2.6}$ \\
    & NT-Xent & $19.5_{\pm0.1}$ & $50.4_{\pm0.3}$ & $65.6_{\pm0.8}$ & $22.2_{\pm0.7}$ & $53.7_{\pm2.0}$ & $69.5_{\pm0.4}$ \\ 
    \bottomrule
  \end{tabular}
\end{table*}
\vspace{-0.05cm}
\subsection{Experimental setups}
Log mel-spectrograms are used as the audio features, which are extracted using a \num{1024}-points Hanning window with \num{320}-points hop size and \num{64} mel bins. All the models are trained for \num{50} epochs using Adam \cite{kingma2014adam}. The learning rate is set to \num{1e-4} or \num{5e-5} and is decayed to \num{1}/\num{10} of itself every \num{20} epochs. The batch size is set to \num{32} for the AudioCaps dataset and \num{24} for the Clotho dataset. We perform experiments by freezing and fine-tuning the pre-trained models. The best model is selected based on the sum of recalls on the validation set.

For all triplet-based losses, the distance margin $m$ is set to \num{0.2}, and the temperature hyper-parameter $\tau$ in the NT-Xent loss is set to \num{0.07}. For triplet-weighted loss, we follow the hyper-parameter settings in \cite{Wei_2020triplet_weight}, that is, $Q=2, \{a_0=0.5, a_1=-0.7, a_2=0.2\}$, and $P=2, \{b_0=0.03, b_1=-0.4, a_2=0.9\}$. The dimension of the shared embedding space is \num{1024}. The learned acoustic semantic embeddings are normalized. All experiments are carried out on a single RTX3090 GPU.

\subsection{Evaluation protocol}
Recall at rank $k$ (R$@k$) is used as the evaluation metric, which is the popular cross-modal retrieval evaluation protocol.  R$@k$ measures the percentage of targets retrieved within the top $k$ ranked results, thus the higher the score, the better the performance. We report R$@1$, R$@5$, and R$@10$. All the experiments are repeated three times with different training seeds and we report the mean and standard deviation of the metrics.

\section{Results}
\subsection{Model performance}
The experimental results are shown in Table~\ref{tab:results}. The baseline model is simple but effective. It can be observed that fine-tuning can substantially improve the model performance and lead to state-of-the-art results on both datasets. In addition, we could observe that the scores on the Clotho dataset are relatively lower than those on the AudioCaps dataset, which is consistent with the results in \cite{koepke2022audioretrieval}. The reasons might be two folds: (1) the size of training data is limited in the Clotho dataset, (2) captions for each audio clip in the Clotho dataset are more diverse thus it is challenging to map the diverse captions close with each other in the shared acoustic semantic embedding space. 

\subsection{Metric learning objectives}
As can be seen in Table~\ref{tab:results}, the NT-Xent loss shows stable performance on both AudioCaps and Clotho datasets regardless of whether the pre-trained encoders are fine-tuned or frozen. It outperforms the other three triplet-based losses in most cases on both text-to-audio and audio-to-text retrieval. For the other three triplet-based losses, the models trained via triplet-sum are not as good as others especially when the encoders are frozen. However, they achieve similar performance on the Clotho dataset when the encoders are fine-tuned. The triplet-max and triplet-weighted losses achieve similar results on the AudioCaps dataset, but the models trained via triplet-weighted loss perform less well than others on the Clotho dataset when the encoders are frozen. The weighting method does not bring substantial improvements. The reason might be that the weighting function has many hyper-parameters to be tuned while we have used the values in the original work \cite{Wei_2020triplet_weight}. Tuning the hyper-parameters for the dataset we evaluated may lead to better performance. Overall, the NT-Xent loss outperforms other three triplet-based losses on both text-to-audio and audio-to-text retrieval in most situations.

In our experiments, we also found that the two losses based on hardest negative mining, the triplet-max and triplet-weighted losses, are sensitive to the training settings, while the other two, triplet-sum and NT-Xent, are more robust and stable to different training settings. For example, triplet-max and triplet-weighted need a good learning rate, otherwise the models are difficult to converge. On the AudioCaps dataset, all the models are trained with a learning rate of \num{1e-4} and all converge well. However, the models trained via triplet-max and triplet-sum with such a learning rate do not converge on the Clotho dataset. Furthermore, triplet-max and triplet-weighted are sensitive to the initialization of model parameters, and we found that some of the models with different training seeds may not converge under the same training settings.

In addition, we study the impact of batch size on the AudioCaps dataset without fine-tuning the encoders. The results are shown in Table~\ref{tab:results_batch}. We can observe the performance of models trained via triplet-max and triplet-weighted losses degrades considerably when the batch size is increased to \num{128}, while models trained via the other two losses (triplet-sum and NT-Xent) show stable performance with respect to the change of the batch size. This is somewhat inconsistent with results in the literature, where it is generally believed that larger batch sizes could lead to better performance for the triplet-based losses \cite{schroff2015triplet}. We found the triplet-max and triplet-weighted do not converge with a batch size of 128. The reason might be that the learning rate should also be tuned to adapt for a larger batch size. Overall, triplet-sum and NT-Xent losses are more robust to different training settings and datasets, while triplet-max and triplet-weighted are more difficult to train.

\section{Conclusions}
We have presented a simple but effective model to learn the Acoustic Semantic Embeddings for the free-form language-based audio-text retrieval task, then we studied the impact of metric learning objectives based on our model in a constant training setting. We empirically demonstrated that the metric learning objectives have a significant impact on the model performance, where the NT-Xent loss outperformed the popular triplet-based losses and showed stable performances with respect to different training settings and datasets. The triplet losses with hard negative mining need careful tuning of hyper-parameters, and are sensitive to the initialization of model parameters.  

\section{Acknowledgements}
This work is partly supported by a Newton Institutional Links Award from the British Council, titled ``Automated Captioning of Image and Audio for Visually and Hearing Impaired" (Grant number 623805725) and grants EP/T019751/1 and EP/V002856/1 from the Engineering and Physical Sciences Research Council (EPSRC). For the purpose of open access, the author has applied a Creative Commons Attribution (CC BY) licence to any Author Accepted Manuscript version arising.

\bibliographystyle{IEEEtran}
\bibliography{mybib}

\end{document}